\documentclass[12pt]{article}
\usepackage[english]{babel}
\usepackage{amssymb,graphicx,wrapfig,hyperref}
\usepackage{amsfonts}
\usepackage{makeidx}

\begin{document}

\title*\begin{center}
{\textbf{{\LARGE 
Hadron interactions in the U-70 energy range:\\ results and problems. }}}
\end{center}

\begin{center}
V. A. Petrov,  N. P. Tkachenko
\end{center}

\begin{center}
A.A.Logunov Institute for High Energy
Physics, \\NRC "Kurchatov Institute", \\Protvino, RF\vspace{-2.1mm}
\end{center}

\begin{center}
Abstract
\end{center}
\textit{In anticipation of the modernization and upgrade of the proton synchrotron at the Institute
of High Energy Physics\footnote{\href{http://web.ihep.su/library/pubs/prep2025/ps/2025-2.pdf}
{Vasiliev A.N. et al. Research directions at the proton accelerator with energy of 400 - 1000 GeV.\\
http://web.ihep.su/library/pubs/prep2025/ps/2025-2.pdf}},
we provide an overview of significant results, their impact on the development of high-energy
physics, and give some new conceptual interpretations of these results.\vspace{-3.1mm}}

\section*{Historical prolegomena\vspace{-2.6mm}}

From late 1967 until the early 1970s, the Soviet proton synchrotron (better known as the "Serpukhov
accelerator" or "U-70"), with a nominal laboratory energy of up to 76 GeV, was the world leader. During
these few years, several  results of fundamental significance were obtained (partially in collaboration
with physicists from CERN, France, and other countries). Let us briefly outline the main provisions of the
paradigm that reigned at that time in the field of strong interaction theory.\vspace{-4.1mm}

\subsection*{ Conceptual atmosphere of the 1960s and early 1970s.\vspace{-2.1mm}}

As for the ideological basis, it largely remained in the nets of the theory of analytic S-matrix with an
eye on the self-consistent "bootstrap" scheme, actively and successfully promoted by G.F. Chew and
his supporters \cite{Ch1} . By the late 1960s and early 1970s, it seemed unshakable that the main
force of hadron interaction at high energies was realized by the Pomeron exchange, a C-even
"vacuum" Reggeon with intercept 1 \cite{Ch2} , so that multi-Pomeron exchanges asymptotically die
out with increasing energy, and, accordingly, the total cross sections of hadron interactions tend to
constant values. A major contribution to the development and substantiation of this scenario (with the
blessing of L.D. Landau) was made by Soviet theoretical physicists I. Ya. Pomeranchuk \cite{Pom} and
V. N. Gribov
\cite{Grib} .

Nevertheless, there was also a "dissident" point of view.
In the early 1950s, W. Heisenberg (the author of the general S-matrix program), while studying
processes at high energies within the framework of the nonlinear scalar theory $ \varphi^{4} $,
concluded that the inelastic cross sections of hadron interactions should increase at ultrahigh
energies as the square of the logarithm of the collision energy \cite{Hei} . As a consequence,
the total cross-sections should also grow, and at least not slower.

Almost immediately this result was subjected to crushing (though not entirely fair) criticism by Landau
et al. and was virtually consigned to oblivion\footnote{It must be said that Heisenberg did not enter
into the controversy, but remained faithful to his result until the mid-1970s, having witnessed its
experimental confirmation.

It is also appropriate to recall here that in the early 1960s Froissart and Martin rigorously proved that
the quadratic-logarithmic growth of total cross sections is the maximum allowable within the framework
of the general principles of the theory.}.

About 20 years later, the possibility of a growing total cross section underwent a renaissance in the
work of T. T. Wu and H. Cheng (1970)\cite{Wu}  when summing ladder diagrams with exchanges
of massive vector mesons. Here the criticism of Landau and Co. was no longer applicable at all.
Nonetheless,  this result did not produce a revolution (just like Heisenberg's result 18 years earlier)
although some (particularly when extracting cross-sections from the cosmic ray data) already tried
to use logarithmically increasing terms\footnote{However, these attempts did not make  much of an
impression due to the low accuracy of the cosmic ray data.} . 

And yet,the overwhelming majority of the physics community continued to believe in the theory of
the "classical" Pomeron with a unit intercept and, as a consequence, with asymptotically constant
total cross sections which seemed natural given the finite range of nuclear forces.\vspace{-5.1mm}

\subsection*{"Serpukhov Effect".\vspace{-2.1mm}}

Measurements of the total cross-sections in $ pp,\bar{p}p, \pi^{\pm}p, K^{\pm} $ -interactions
carried out with unprecedented accuracy at the U-70 by Yu.D. Prokoshkin's group (with the
participation of  J.V. Allaby(CERN) and G. Giacomelli (INFN)) \cite{Prok1}   attracted widespread
attention and, as can be seen in Fig.1, seemed to indicate a seemingly evident transition to an
asymptotic regime of the total cross sections as functions of energy reaching constant values,
thereby confirming the prevailing paradigm under the sign of the "classical Pomeron."

\begin{wrapfigure}[26]{l}{64mm}
\vspace{-5.1mm}
~\hspace{-8.6mm} \includegraphics[width=76mm]{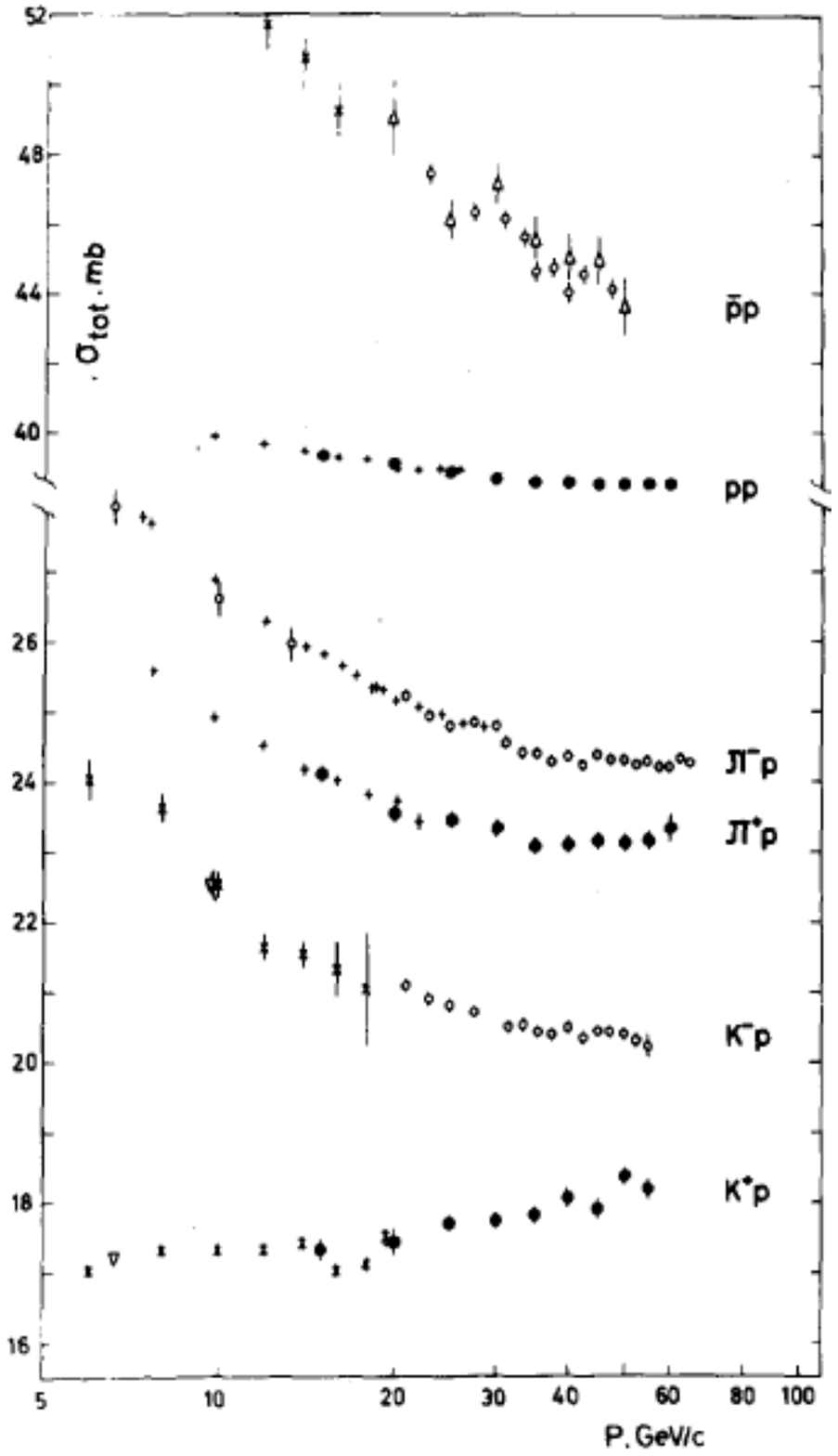}\vspace{-3.1mm}
\caption{Results of the total cross-section measurements as they were at the end of 1971, taking into account the latest data from the U-70.}
\label{Fig2Denisov}
\end{wrapfigure}

An outlying \textit{increase} of the cross section of the $K^{+}p$ interaction was interpreted as a
transient type of approach to a constant cross section from below, in contrast to, say, the case of
$pp$. Here is the suggestive conclusion of the paper \cite{Prok1}:
"\textit{... the total cross-section for} $K^{+}p$ \textit{will approach the asymptotic value from
below...".}
The general trend of these experimental findings marking the transition to a new regime of energy
evolution has been dubbed "Serpukhov effect."\vspace{-5.1mm}

\subsection*{Scaling in Inclusive Processes.\vspace{-2.1mm}}

Note that in the energy region under discussion, the elastic cross sections, unlike the total ones,
continued to \textit{decrease}. In such circumstances the natural reason for the observed constancy
or a growth of the total cross-sections\footnote{Let us remember that $\sigma_{tot} =
\sigma_{elastic} + \sigma_{inelastic} $.}  is the continuous increase in \textit{inelastic} cross
sections, including the cross sections of multiple production processes (partly due to the rapid
increase in opening channels).

In 1967, A. A. Logunov \cite{Log}  proposed a new method for describing multiple processes in
terms of probability densities of detected particles. Somewhat later, this type of cross-sections and sets
of processes was dubbed "inclusive." Experiments, however, used a slightly different type of
cross-section (taking into account multiplicity in the relevant channels), which represented the density
of the average number of detected hadrons.\footnote{The isomorphism between these types of
inclusive cross-sections was established in \cite{Ezh}. } This type of distribution, however, has long
been routinely used in the study of cosmic rays.

The group led by Yu.D. Prokoshkin  has found (in collaboration with a team from CERN)\cite{Prok2}\
that the ratios of momentum distributions (inclusive spectra) of secondary hadrons cease to depend
on energy, remaining functions only of the ratio of the momentum of the detected hadron to the
maximum possible. So the proportional scale change of the momentum of the detected hadron and
the maximum momentum (of the order of the momentum of the initial particles) leaves the indicated
ratios unchanged.

Just a few months later, a group of theorists led by a Nobel laureate Yang-Chen Ning, building on
the results obtained on the U-70, formulated the so-called "limiting fragmentation hypothesis," which
provided a physical justification for "scale invariance."

A little while later, another Nobel prize winner, R. Feynman, spoke on this topic, linking “scale
invariance” with the properties of the “classical” Pomeron. This was qualified in the literature as
“Feynman scaling”. It was subsequently found that the "limiting fragmentation" effect is confirmed
up to the highest energies, while "Feynman scaling" is violated for particles slow in the
center-of-mass frame but is preserved in the fragmentation regions. It is worth noting that the
phenomenon discovered at the U-70 supports “limiting fragmentation” albeit the cause of the
phenomenon remains unclear to this day.

As often happens, it was later discovered that scaling had been predicted by Heisenberg back in
1963\cite{Hei2}.

\subsection*{Quest for quarks.\vspace{-2.1mm}}

In early 1964, two papers called into question the entire S-matrix paradigm, according to which in
the world of hadrons - nucleons, pions, kaons, hyperons... - a "nuclear democracy" reigns and there
are no more elementary units: each hadron is no more and no less elementary than any other.
The new creed was proclaimed by M. Gell-Mann and G. Zweig\footnote{Much later it turned out that
even earlier, A. Petermann had proclaimed the same idea, but for some still unclear reasons, the
publication of his article, sent earlier than Gell-Mann’s article, was delayed for more than a year.} .
Although Gell-Mann himself was initially unsure of the actual existence of three new elements — quarks
— the question of detecting them in hadronic reactions arose almost immediately. This was the goal of
the Protvino IHEP group, led by L.G. Landsberg, set for themselves at the U-70 accelerator. 

Since quark charges do not exceed
two-thirds of the minimum charge observed so far, the hope was to observe quarks among very weakly
ionizing tracks. Moreover, the U-70 made it possible to detect quarks with masses up to
$ \approx 5~\mbox{GeV} /c^{2} $,  an enormous value for that time.
The result was negative\cite{QU} , but given that other searches for quarks had already been conducted
for nearly five years prior, the question arose: do they even exist in free form? This question remains
unanswered and is known as the "confinement problem," the supposed solution of which states that the
physical hadronic spectrum does not directly correspond to the system of elementary gauge fields.

It is appropriate here to compare this “negative experiment”at the U-70  with the famous
Michelson-Morley experiment (1887) to search for the “ether wind”, the negative result of which
became a powerful argument in favour of the theory of relativity.

So, we've commented on some of the key achievements obtained in experiments at the U-70 accelerator
and the theoretical conclusions drawn from them. 

\section*{To sum it up...}

Let's summarize the results obtained at U-70:\vspace{-3.1mm}

\begin{enumerate}
\item In the U-70 energy range, we observe a transition from the dominance of secondary trajectories
          (decreasing cross sections) to a "flattening" regime ("Serpukhov effect"), which, according to the
          prevailing views at the time, would have meant the dominance of the "classical Pomeron." A year
          later the first data from the NAL accelerator \cite{Bar}  seemingly confirmed this trend ("within...
          errors the total cross section remains constant in the energy interval from 48 to 196 GeV".)
\item A potentially alarming  increase in the total cross-section in the $K^{+}p$ channel
          was detected, but no great significance was payed to it.\vspace{-3.1mm}
\item Inclusive cross sections in hadron interactions revealed a universal dependence on the
          collision energy (if any) for fixed values of the scale-invariant variable. So their ratios for different
           species of detected hadrons do not depend on the initial energy.\vspace{-3.1mm}
\item U-70 showed that even up to large masses of the order of $ 5 \;GeV/c^{2} $, quarks
          are not produced in a free state. This was one of the powerful arguments for promotion of the
         "confinement" hypothesis.\vspace{-2.1mm}
\end{enumerate}

What happened next? The paper that published the phenomenon of $ K^{+}p $ total cross-section
growth appeared on September 20, 1971.

The official inauguration of CERN's Intersecting Storage Rings (ISR), the world's first proton-proton
collider, took place on October 16, 1971, in Geneva, Switzerland. A symbolic episode: the opening
speech was given by one of the founders of CERN and the theorist who first predicted the increase
of the total cross-sections of nucleons at high energy, Werner Heisenberg.

A little over a year later, the rise was found in two experiments at the ISR\cite{Am}.

Nonetheless, even after this, the most ardent supporters of the {\it classical Pomeron} with
$\alpha_{ \mathcal{P} } (0) = 1$ fought a rearguard action for almost two years, trying to save their
creed. 
Some even abandoned this area of particle physics altogether.

As to the underlying physical cause and significance of the supposedly unlimited growth of cross sections,
if we set aside the diversity of unexplained parameters in a couple of dozen models, they remain on the
agenda.

In the next section, we'll attempt — in the style of
"alternative history"— to understand whether, while remaining within the theoretical framework of the
1950s and 1960s, it would have been possible to arrive at different conclusions based on the same
experimental results.

\section*{" If I had it to do all over again..."\footnote{from the Roy Clark hit}. }

Let's cast our minds back to the autumn of 1971. We have new data on total cross sections, from the
cutting edge of modern high-energy physics, and we're trying to understand what they do point to.
The theoretical mainstream assures us that we're at the beginning of a regime of asymptotic constancy
of total cross sections, and this signals the undeniable triumph of the Regge-Pomeron paradigm.
Accordingly, we would take the following as a tentative expression for fitting which embodies the
property of the "classical" Pomeron, $\alpha_{P}(0)= 1$ :
\begin{equation}
\sigma_{tot}^{AB}(s)= C^{AB}- D^{AB}/[\ln s+ E^{AB}]) 
\end{equation}
for $ AB = pp, K^{+}p $
where the second (logarithmic) term in brackets takes into account the contribution of 2-Pomeron
exchange. The absence of contributions from the secondary Reggeons follows from their degeneracy
due to duality arguments \cite{Ros}.

For the rest of channels ($ \bar{p}p,\pi^{\pm} p, K^{-}p $ ) we have\vspace{-1.6mm}
\begin{equation}
\sigma_{tot}^{AB}(s)= C^{AB}- D^{AB}/[\ln s+ E^{AB}]
+R^{AB} /s^{\eta}
\end{equation}
with 
\begin{equation}
R^{\bar{p}p}>0,R^{K^{-}p}> 0; R^{\pi^{-}p}= - 2 R^{\pi^{+}} > 0. 
\end{equation}

The secondary Reggeon intercept is $\alpha_{R} (0)= 1-\eta, \eta < 1/2$.
The comparison of these formulae with the U-70 data is shown in Fig.1 of the Appendix. We see
that the quality of the fit is quite poor.

So we are now to take seriously the alternative possibilities outlined in the Heisenberg (1952) and
Cheng-Wu(1970) writings.  The tentative parametrization is Eqs.(1)-(2) plus the terms
\begin{equation}
L^{AB}\ln^{2}s, L^{AB}> 0.
\end{equation}
Here we just add the logarithmic term accounting for the work of Heisenberg and Cheng-Wu.

In this way we could predict, in particular, the values of the total cross section in $ pp $ interactions
at the ISR energies that were about to be measured at CERN according to two options: the classical
Pomeron and the growing Heisenberg-Cheng-Wu variant.

Fig. 2 shows that the Heisenberg-Cheng-Wu-style parametrization implies a statistically justified
non-zero coefficient $ L^{AB} $. So, the end of our hypothetical article from the fall of 1971 could
have ended like this:

\newpage

\textit{"Thus, our description of the U-70 data gives us grounds to assert that}\vspace{-2.6mm}

\begin{enumerate}
\item \textit{The growth of the $K^{+}p $ cross-section is not a transient phenomenon, but will
          continue at higher energies, which can be observed at FNAL energies.}\vspace{-2.6mm}
\item \textit{The "Serpukhov effect" is not the beginning of asymptotically flat behaviour of the cross
          sections, but should be replaced by an increase, which should be observed in the upcoming
          experiments at the ISR at CERN."}\vspace{-2.6mm}
\end{enumerate}

Unfortunately, such a prediction was never made, largely due to the unquestionable authority of the classical
Pomeron theory's supporters. A curious illustration: in March 1973, Ugo Amaldi, the spokesman of one of
two ISR experiments that observed the increase in total cross sections, gave a talk on this to a large
audience in CERN's Great Auditorium. After his talk, he heard the highly skeptical opinion of two leading
physicists—experimentalist Carlo Rubbia (later a Nobel laureate and CERN Director-General) and
renowned theorist Daniele Amati: "Ugo, you're wrong, because the Pomeron intercept is 1!" 

This
example is quite instructive.\vspace{-6.1mm}

\section*{"There is still plenty of good music to be written in C
                major\footnote{A. Sch\"{o}nberg}” \vspace{-2.1mm}}
                
   The last half-century in particle physics has been marked by a widespread and all-consuming desire to discover "new physics." In our view, this desire, largely fueled by the inglorious saga of the "Theory of Everything," has given rise to a false division of our field into "new" and "old" physics. It would be highly imprudent to label any phenomenon, even one that has long eluded experimental discovery and/or explanation, with the dismissive, marginal label of "old physics."
                                                    
With this understanding, the moments of real or alternative history described above inevitably make us think about the  future (hopefully , not too remote).

To this end we would now like to briefly outline the promising (in our opinion) measurements that could be performed at the U-70
(upgraded or not) and which could lead to significant progress in our understanding of the hadronic world.\vspace{-3.1mm}

\begin{enumerate}
\item Some time ago, a proposal for an experiment to precisely measure differential cross sections in pp
scattering at energies of about 50 GeV was published in \cite{Den} to clarify the issue of cross section
oscillations in the very low transferred momenta  region.\vspace{-2.1mm}
\item The work of A. Zaitsev's group at the VES facility on the observation and analysis of neutral
systems $X^{0}$ in the process\vspace{-3.1mm}
 $$
 \pi^{-}+ p\rightarrow X^{0}+n\vspace{-2.1mm}
 $$
and also the experiment OKA in the V. Obraztsov's group with $K^{+}$ beams at U-70 seem very
promising for the search for glueballs\footnote{It would be unfair not to mention the early attempts
of Yu.D. Prokoshkin, who persistently and enthusiastically searched for glueballs from the mid-1980s
to the end of the 1990s both at U-70 and at CERN (GAMS collaboration).}  and light multiquarks.
Several considerations on this matter are expressed in \cite{Ger}.\vspace{-2.1mm}
\item "Last but not least".\vspace{-2.1mm}
\end{enumerate}
 
 Finally, we thought it appropriate to present a few modest ideas about a possible way to upgrade the U-70.
 They are based on works that were once published \cite{Tka} but, for obvious reasons, were not used.
 
In the U-70 accelerator, a proton energy of 70 GeV is achieved with a dipole magnetic field of
$B = 1.2$ Tesla. As is well known, the energy of a synchrotron is proportional to the magnetic field and
radius. If superconducting magnets (SPM) with a field of 10 T are used in the accelerator, then in the
same U-70 tunnel, a proton energy of ~583 GeV can be achieved. Experience in creating SPMs at IHEP
allows us to confidently say that such magnets based on NbTi superconducting cable are entirely
feasible. Moreover, if a slightly modified superconductor composition is used, namely, the same
NbTiTa-based superconducting cable, then a field of up to 12.5 Tesla can be achieved in dipole magnets,
allowing the proton energy in the same U-70 accelerator ring to be increased to 730 GeV. The stated
magnetic fields can be achieved at a cooling temperature of [superfluid] helium of $T=1.8$ K. Such
temperatures have long been achieved at our institute (they were first achieved on a $1m^{3}$
cryostat back in 1984). Conducting such research allows us to restore and develop new young
engineering talent, which is currently virtually lost, as well as lost technological potential in the field
of superconductivity, and, based on this, revive high-field accelerator magnetic technologies.
Further development of such research will allow us to create colliding $pp$ beams at
$\sqrt{s}=1.0\div 1.4$ TeV using extracted $500\div 700$ GeV  $pp$ beams, as was achieved
at SLAC. 

Our point is that even the past results from the U-70 still contain much food for thought. Of course, new
research in this area of energies (and perhaps even higher, but not too much!) could only be
welcomed, as promising new breakthroughs in our understanding of the structure
of Nature.\vspace{-6.1mm}

\section*{Acknowledgements\vspace{-2.1mm}}

We are grateful to the Organizers of the conference "Particle Physics at Medium and High Energies" (2-5 June 2026, Protvino, RF) for an opportunity to give the talk and to its
participants for interesting discussions.
V.P. is thankful to Prof.U. Amaldi for covering moments from the history of the ISR.



\newpage

\noindent {\Large {\bf Appendix}}

\section*{U-70 Experimental Data and Their Predictability}

For this review, we used experimental data obtained at the IHEP at the U-70 accelerator in Protvino
with energies available only at the U-70 accelerator, taken from the PDG database. These include
 $p({\bar p})p$ ($4.5 \lesssim \sqrt{s} \lesssim 12$ GeV) and $\pi^{\pm}p$, $K^{\pm}p$
($4.5 \lesssim \sqrt{s} \lesssim 10$ GeV). Such an energy range was used because the proton beam
hitting the accelerator's fixed target had a range $10 \lesssim p_{lab} \lesssim 70$ GeV/c, while the
pion and kaon beams had a range $10 \lesssim p_{lab} \lesssim 50$ GeV/c. We also retained a very
small number of data points from later observations with cosmic rays. We did not discard them for
data completeness. While they have very large errors and contribute virtually nothing to the statistical
processing of the experimental data they do not contradict the U-70 experimental data.

In the graphs below, we also show experimental data points obtained later at the ISR and Tevatron
accelerators. We do this to compare the theoretical predictions following from the U-70 experimental
data with those obtained at later accelerators. The fitting was performed pairwise in reactions
involving incident particles and antiparticles on the proton target, using the generally accepted
model (see above) of the time, both separately and jointly.

We also considered a modification of this model by adding a “Heisenberg” term to it.

An analysis of the graphs presented, showing their excellent reliability, provides no reason to discard
the growing (“Heisenberg”) variant. The fact that the important parameters of the models differ
significantly when analyzing the experimental data available at the time only speaks to the
inadequacy of the theories on which these models were based. Nothing more.

Apparently, the variant with an indefinitely rising term arose because all experimental data at the
maximum energies of the time demonstrated a decline in the total (and elastic) cross sections. Only
the $K^{+}p$ scattering data, even visually, indicated their possible increase which was regarded
at the time as a transient phenomenon. Moreover, the standard variant (with no indefinite rise) also
indicated a rising possibility, with a subsequent flattening of the cross sections, a fact the mainstream
international community had no doubt about at the time. As further developments in accelerator and
experimental technology have shown, only the inclusion of the indefinite rise in the theory allows for
a correct description of modern experimental data on total and elastic scattering cross sections and,
most importantly, completely eliminates the possibility of their flattening at high energies.

Of course, it is possible to construct a theory of unlimited growth of total and elastic cross sections
different from a simple adding of the logarithmically growing term to the full amplitude (as it takes
place in the eikonal models where the logarithmic growth reveals itself only at PeV-scale energies)
but this simple mechanism works phenomenologically very well with modern data.

%
%
%
%
%
%
%

\newpage

\begin{figure}[h]
~\vspace{-8.6mm}\\
\begin{minipage}[h]{113mm}
\includegraphics[width=113mm]{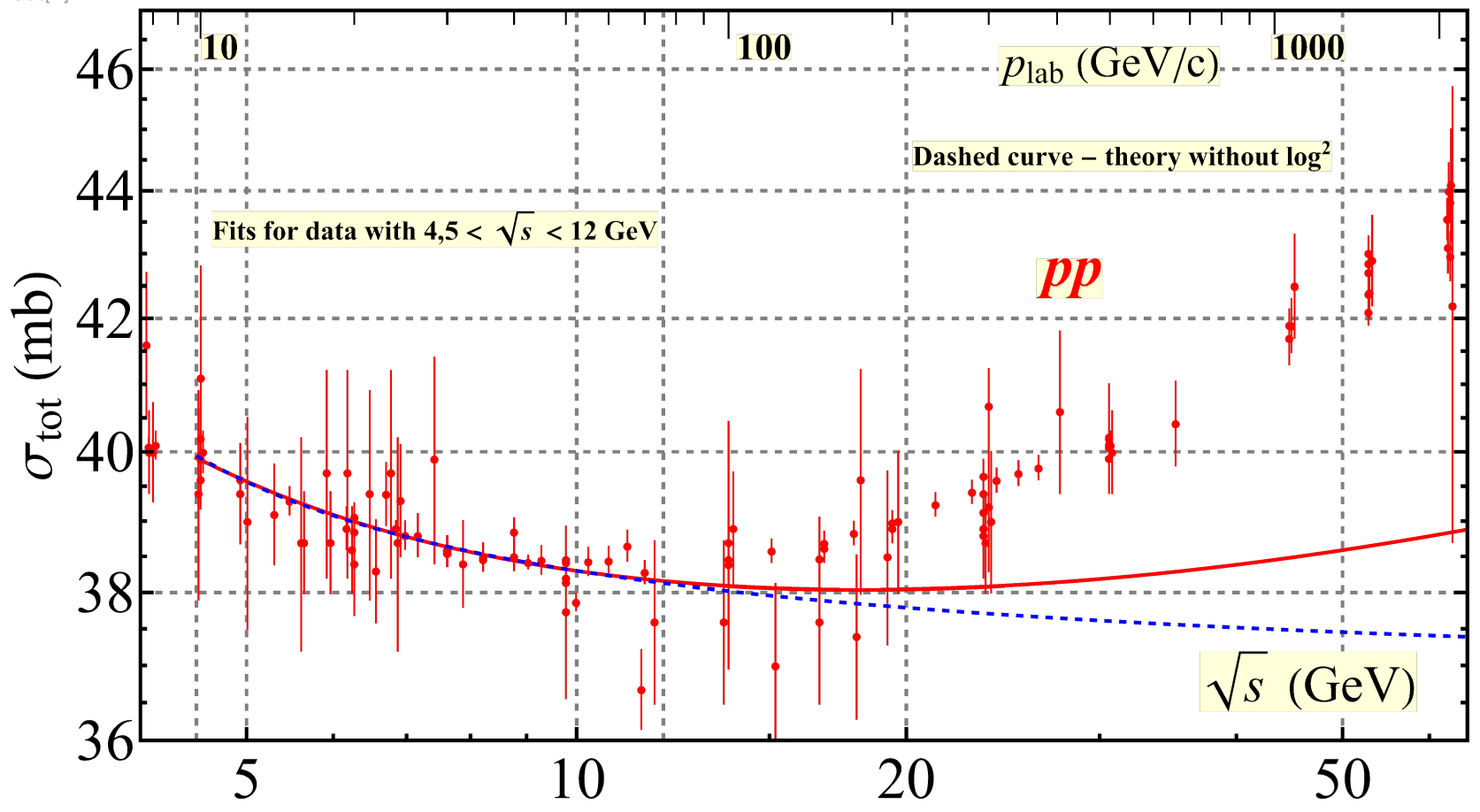}~
\end{minipage}
\hfill
\begin{minipage}[h]{58mm}
\includegraphics[width=58mm]{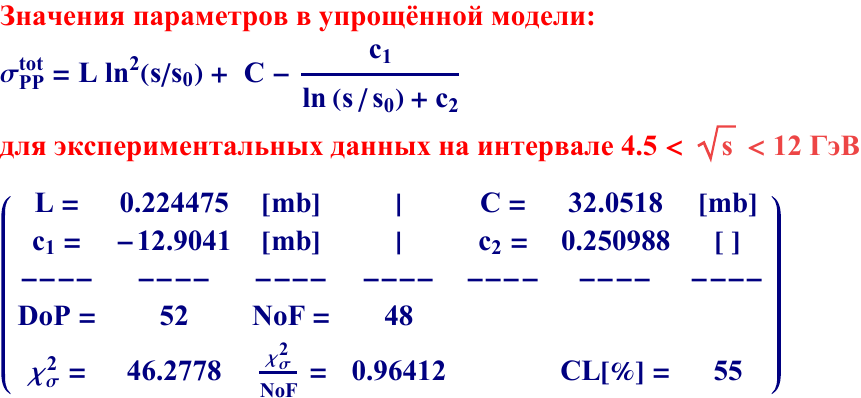}\\
\includegraphics[width=58mm]{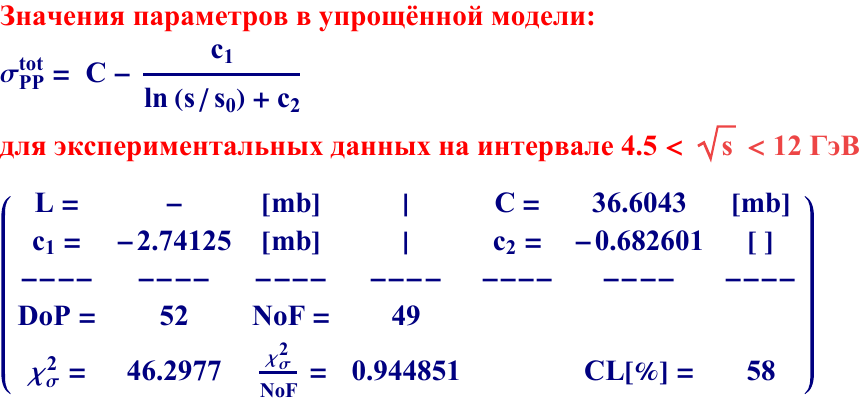}\\
\end{minipage}
\begin{minipage}[h]{113mm}
\includegraphics[width=113mm]{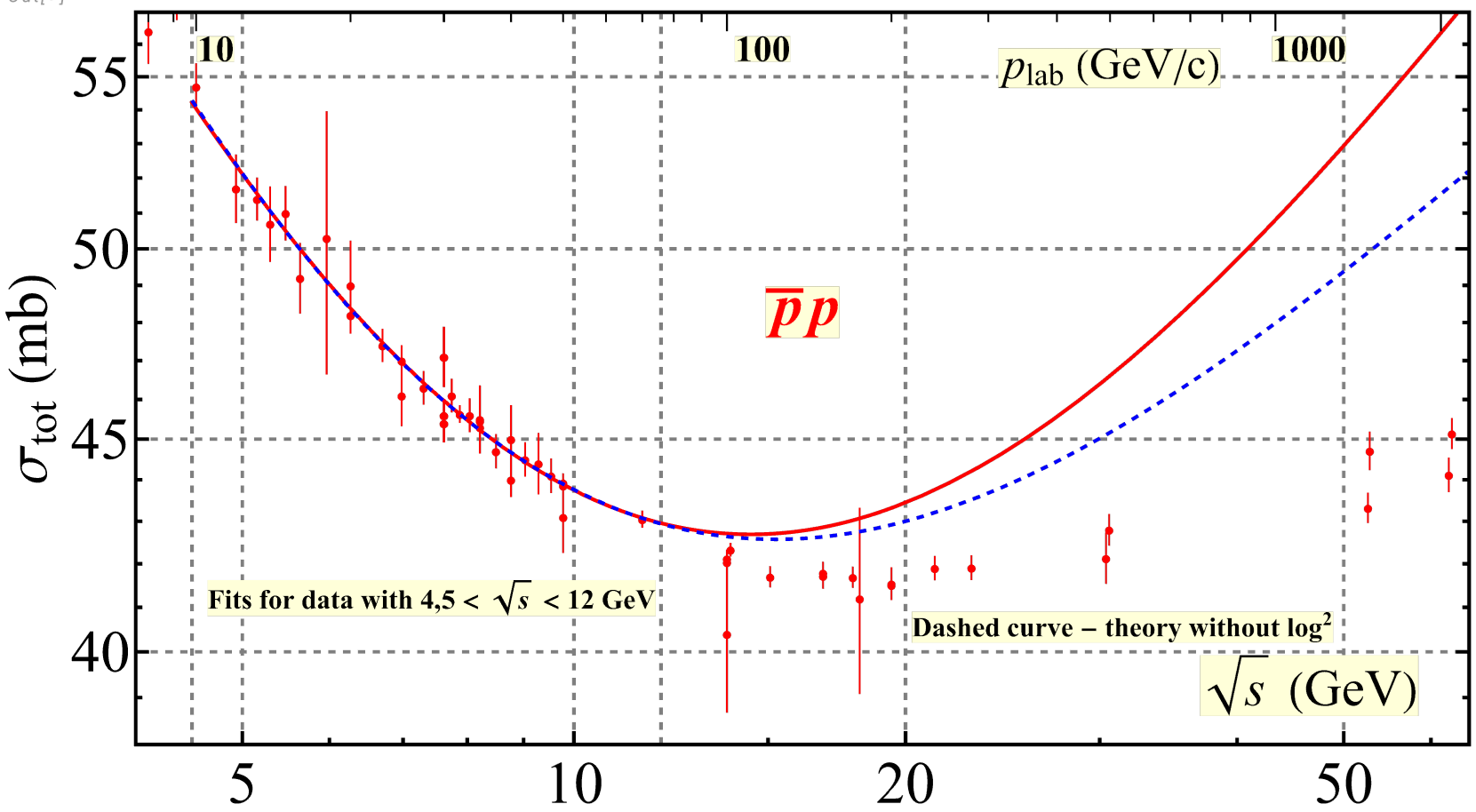}~
\end{minipage}
\hfill
\begin{minipage}[h]{57mm}
\includegraphics[width=57mm]{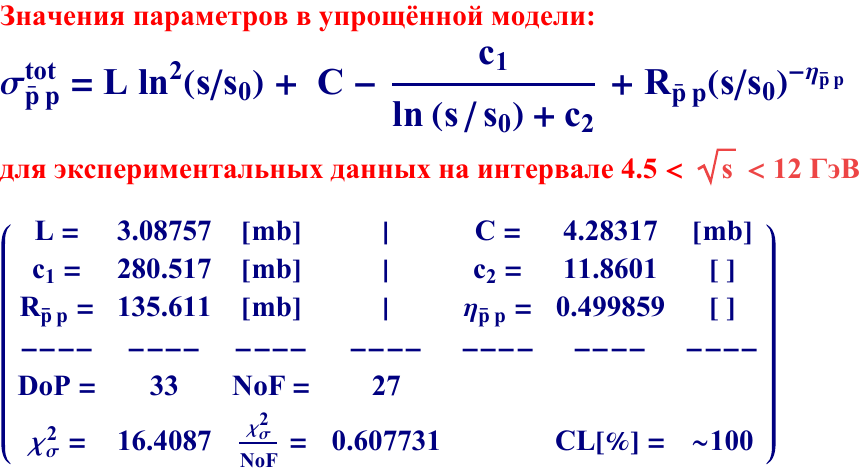}\\
\includegraphics[width=57mm]{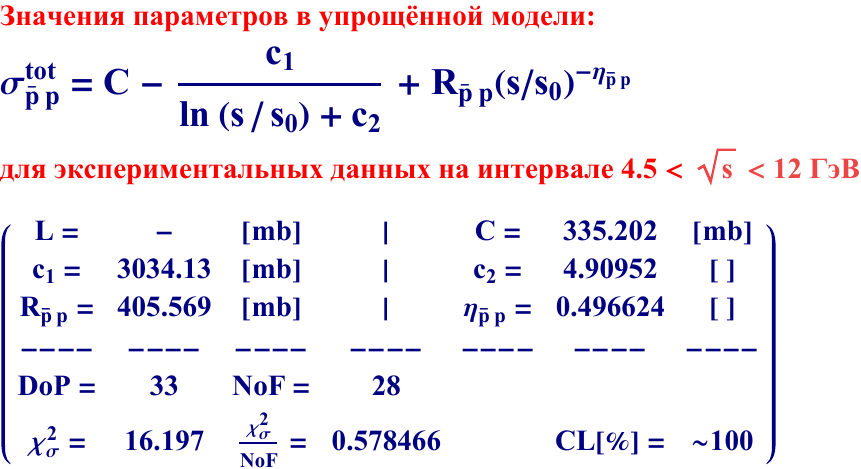}
\end{minipage}
\includegraphics[width=113mm]{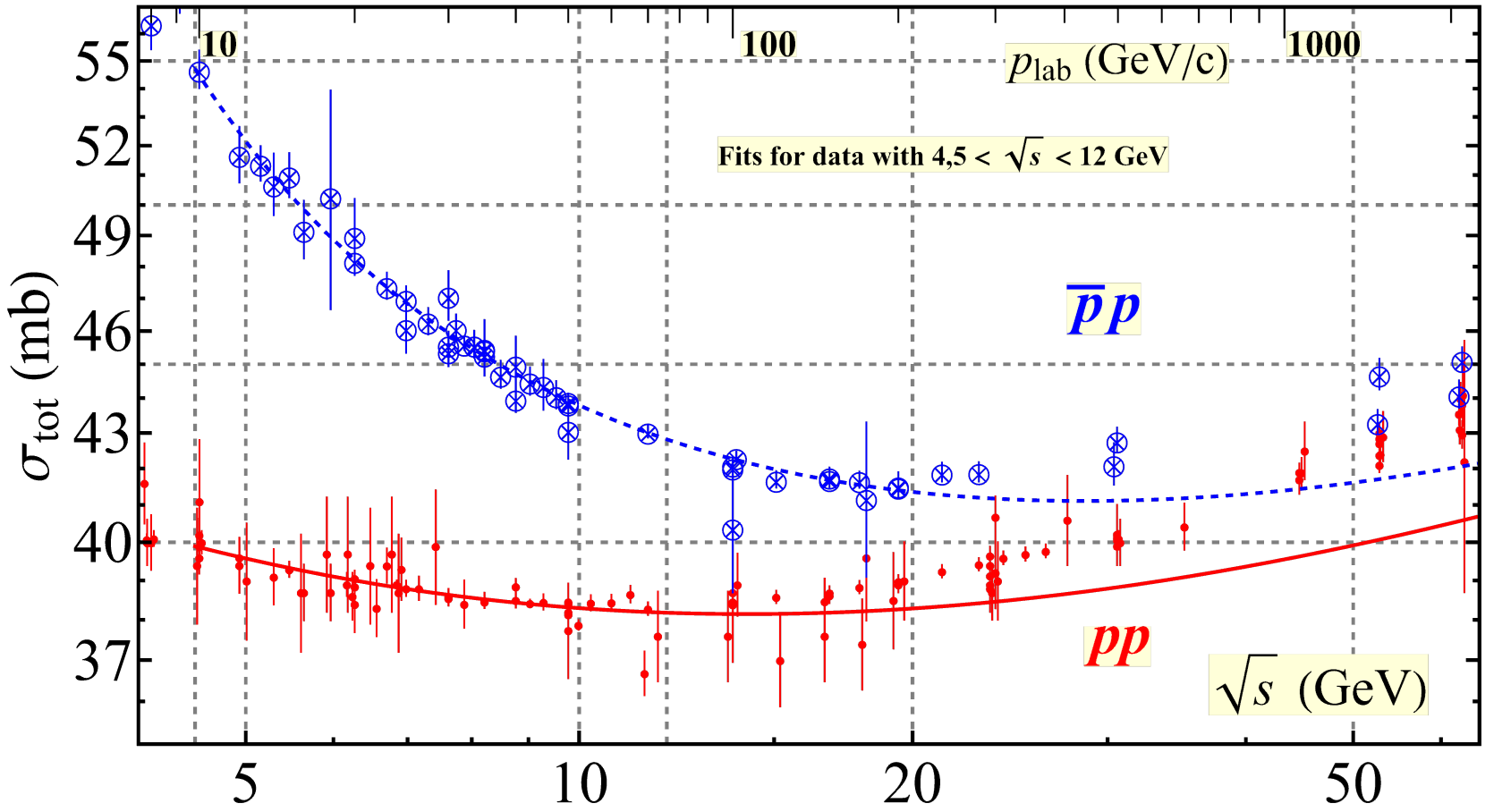} \hfill
\includegraphics[width=59mm]{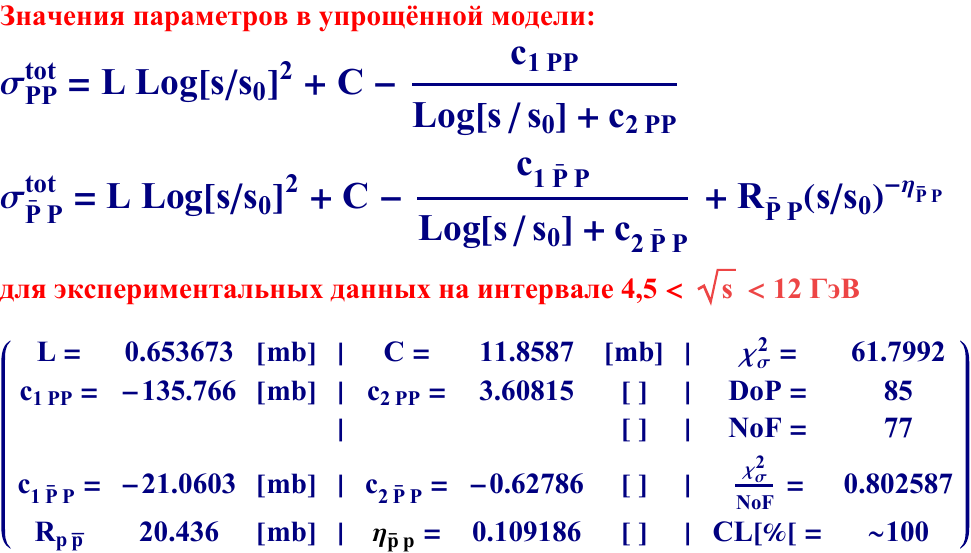}\\
\includegraphics[width=113mm]{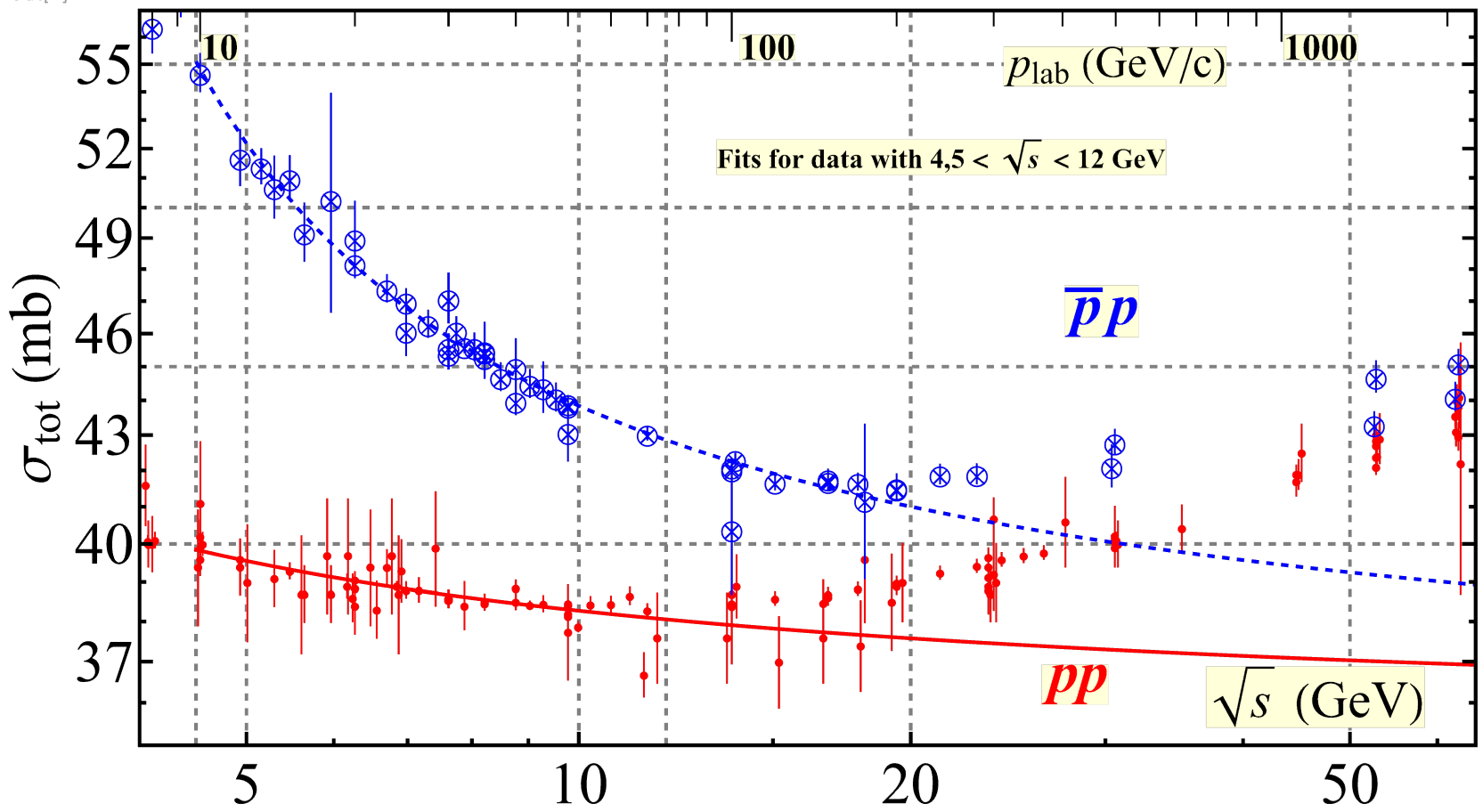} \hfill
\includegraphics[width=59mm]{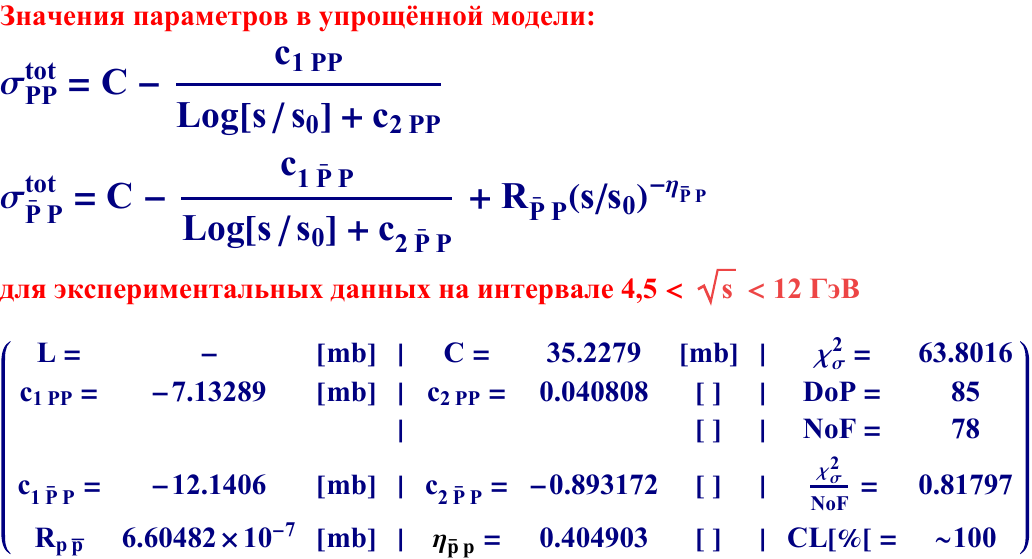}\\
\vspace{-7.1mm}
\caption{$pp$ and ${\bar p}p$ reactions and its parameters}
\label{ppPic}
\end{figure}

\newpage

\begin{figure}[h]
~\vspace{-8.1mm}\\
\begin{minipage}[h]{113mm}
\includegraphics[width=113mm]{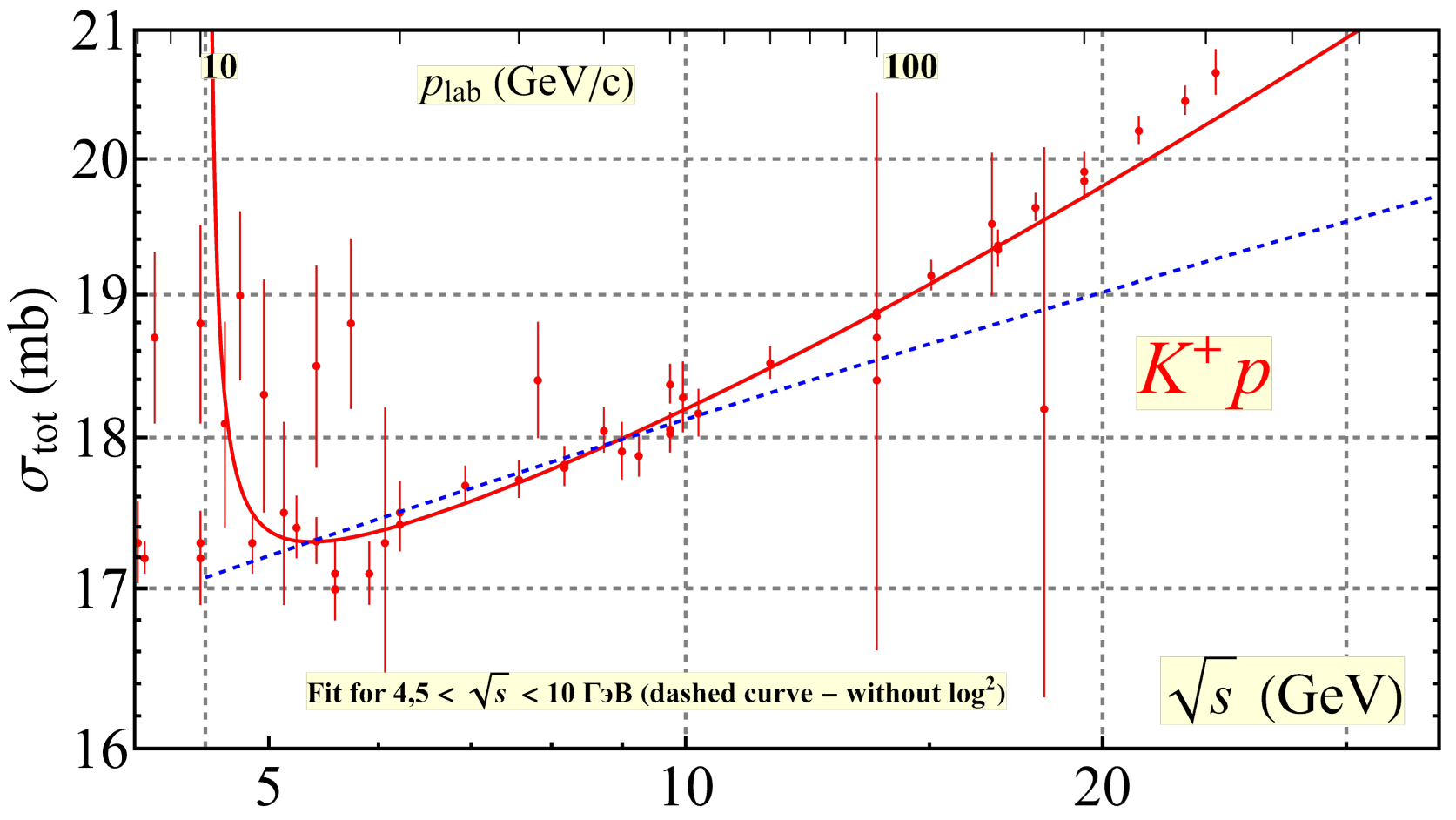}
\end{minipage}
\hfill
\begin{minipage}[h]{60mm}
\includegraphics[width=60mm]{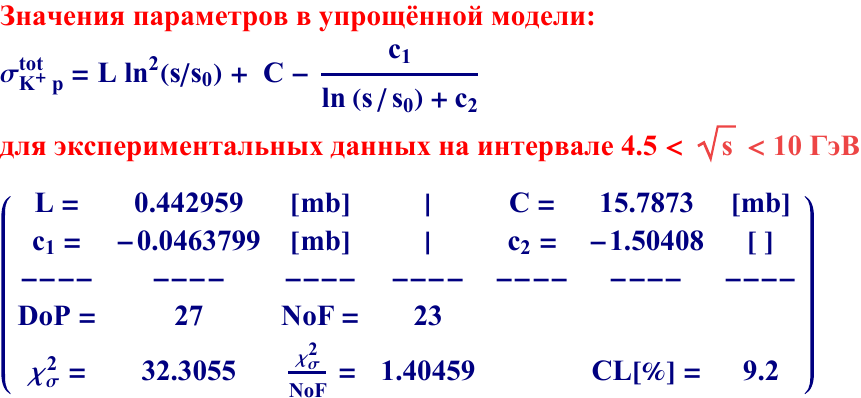}\\
\includegraphics[width=60mm]{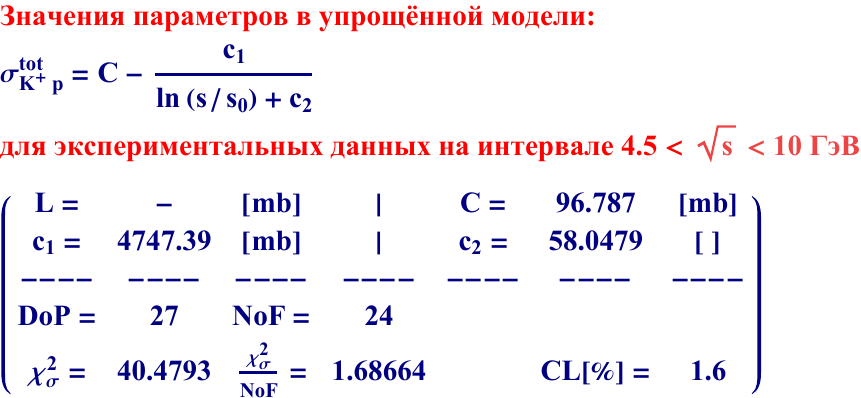}
\end{minipage}
\begin{minipage}[h]{113mm}
\includegraphics[width=113mm]{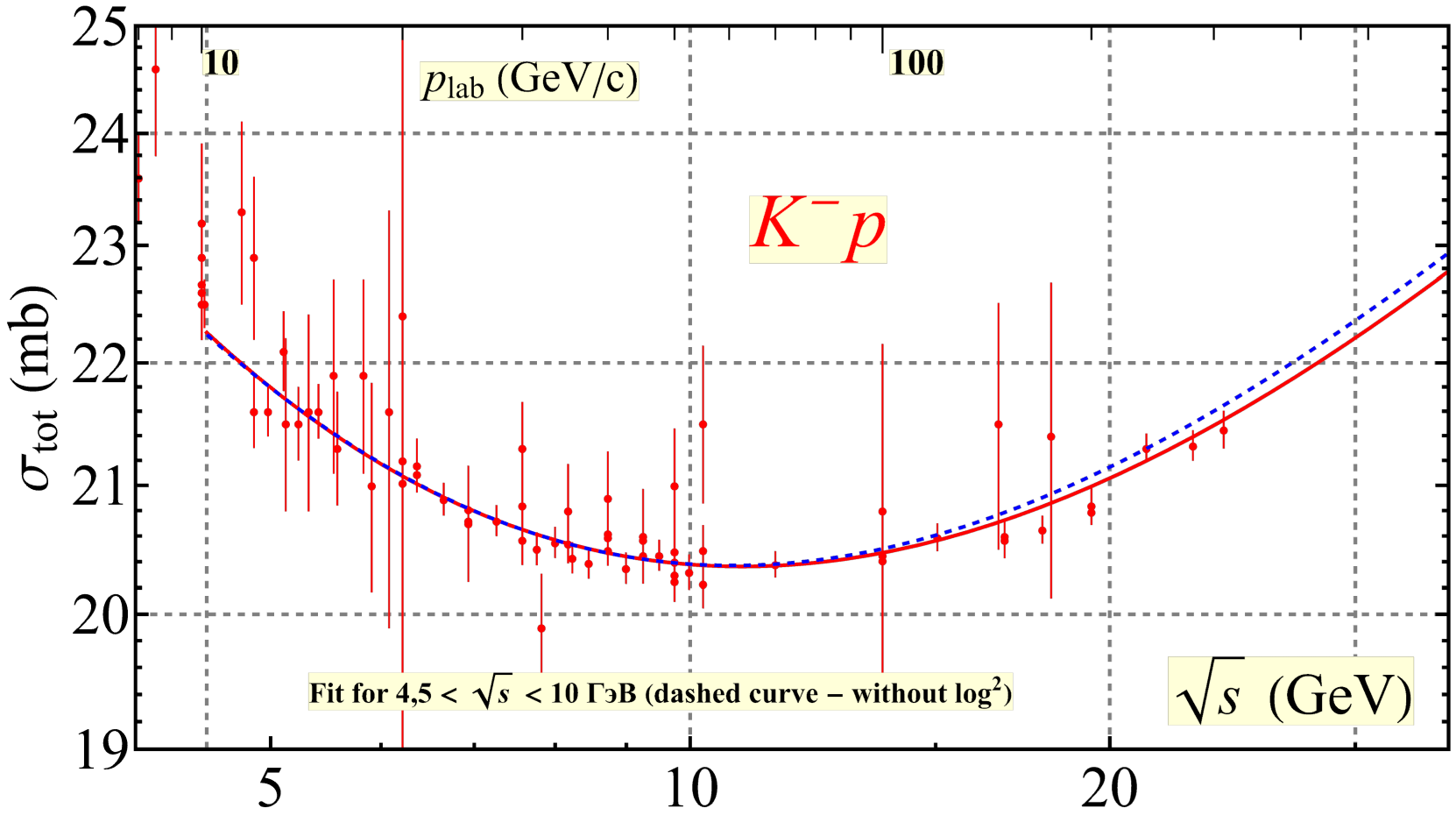}
\end{minipage}
\hfill
\begin{minipage}[h]{60mm}
\includegraphics[width=60mm]{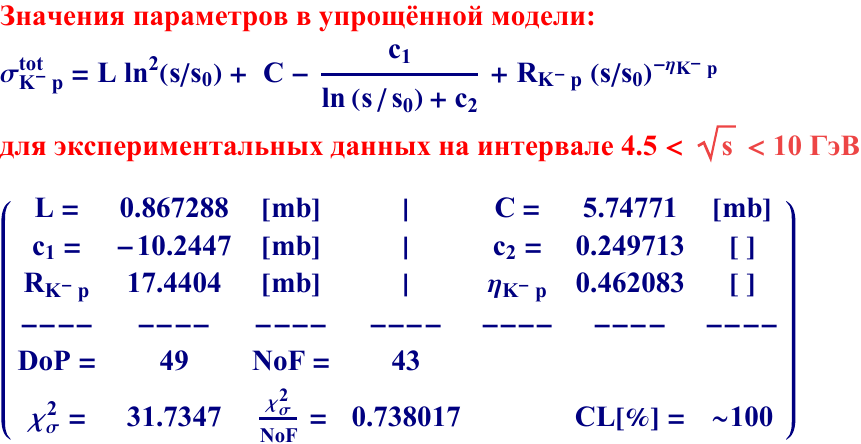}\\
\includegraphics[width=60mm]{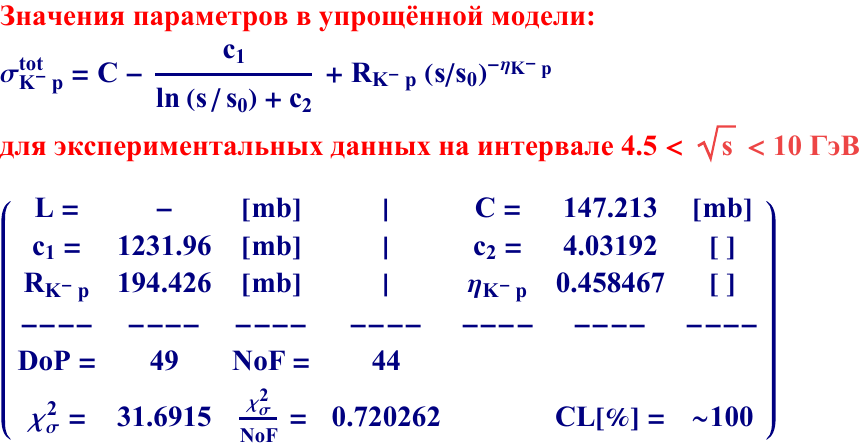}
\end{minipage}
\includegraphics[width=113mm]{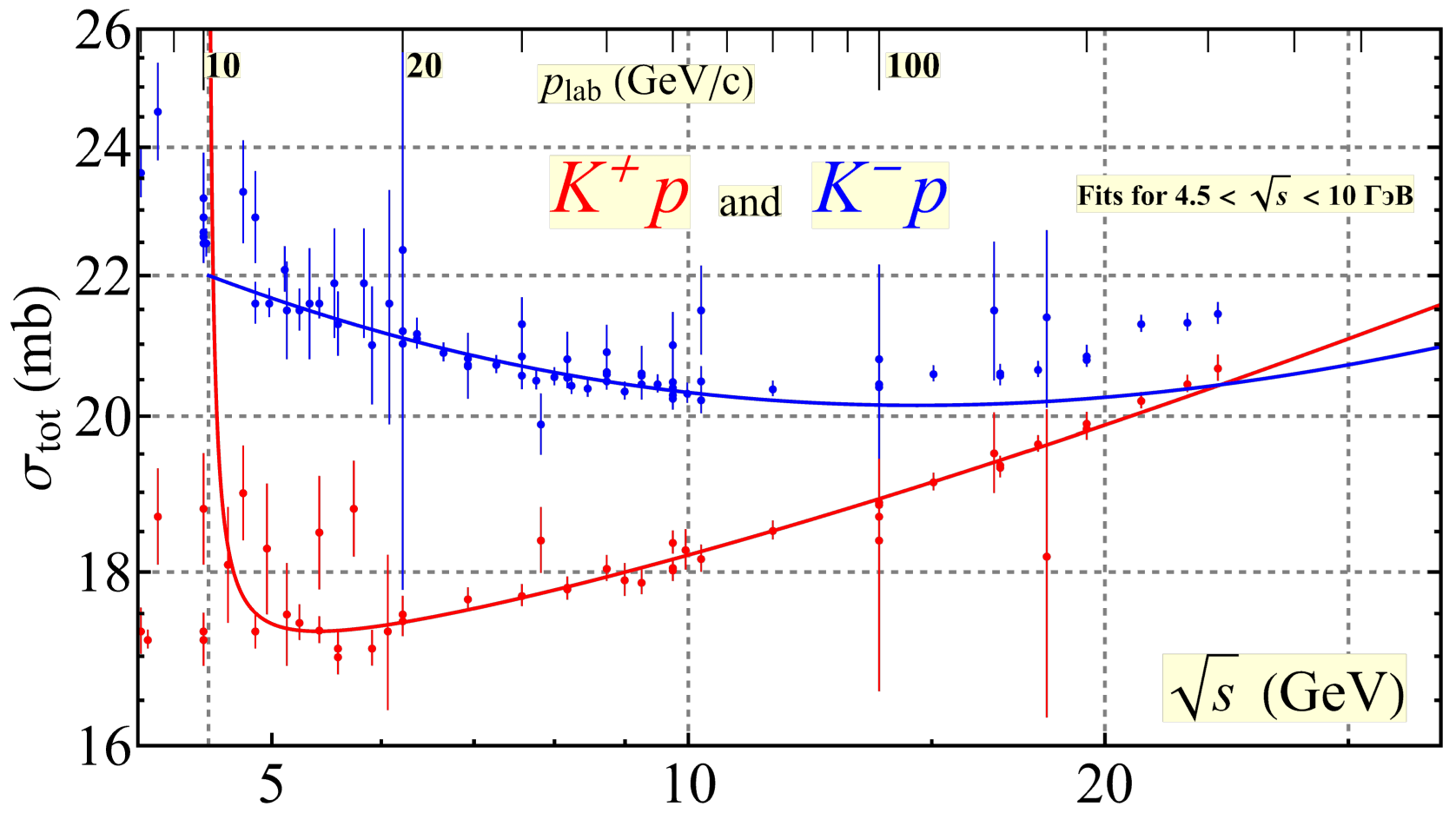} \hfill
\includegraphics[width=60mm]{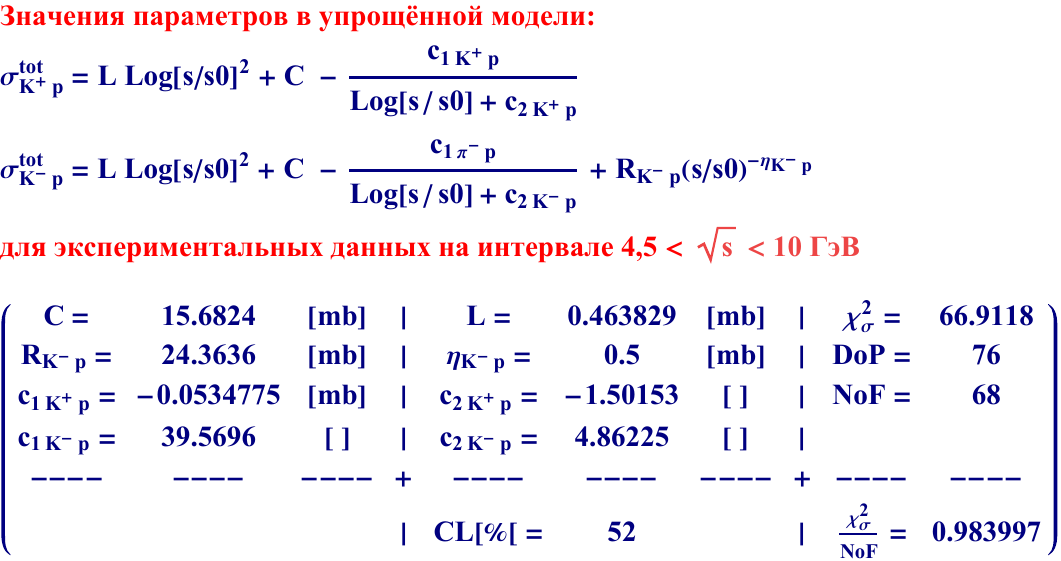}\\
\includegraphics[width=113mm]{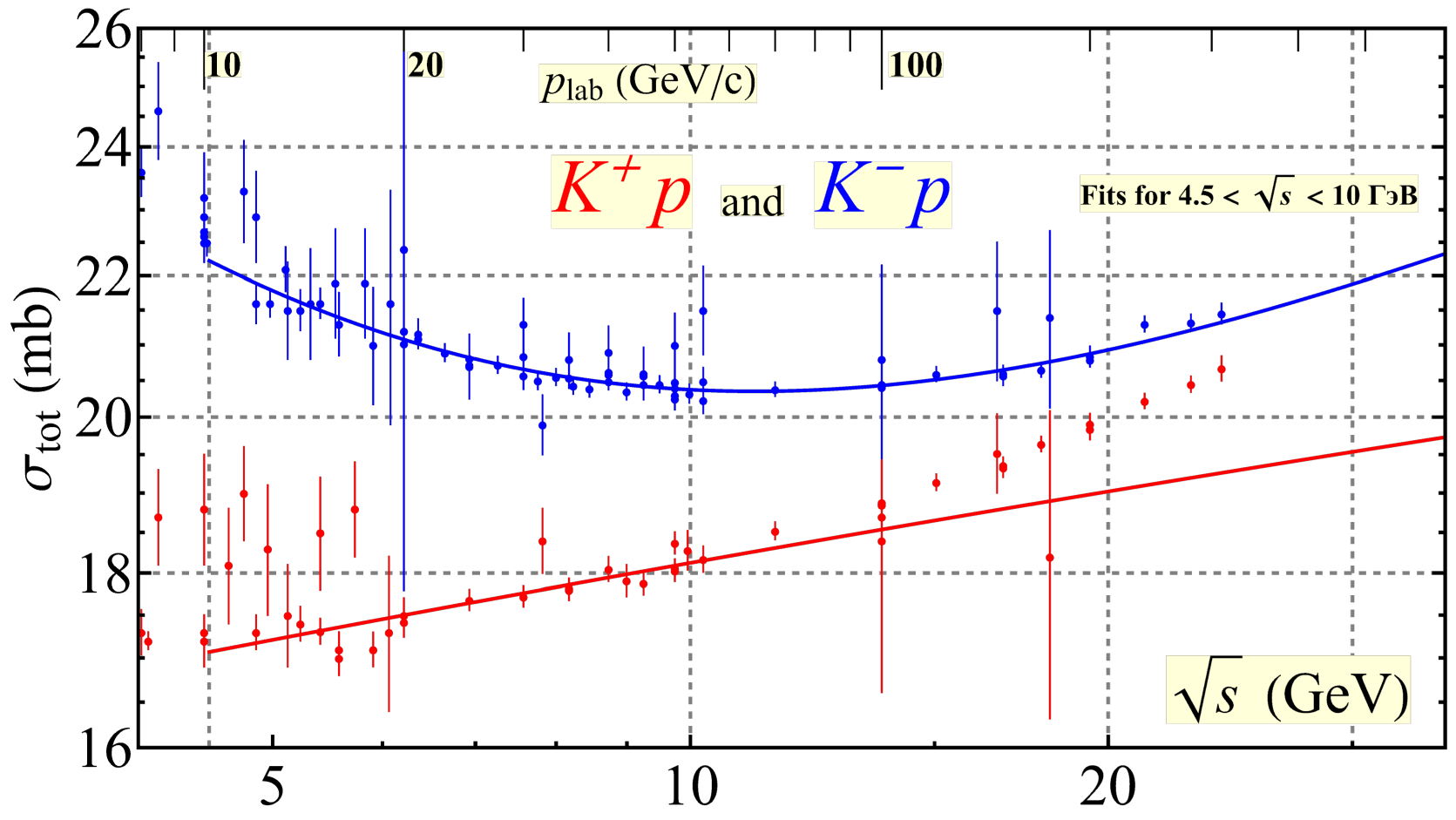} \hfill
\includegraphics[width=60mm]{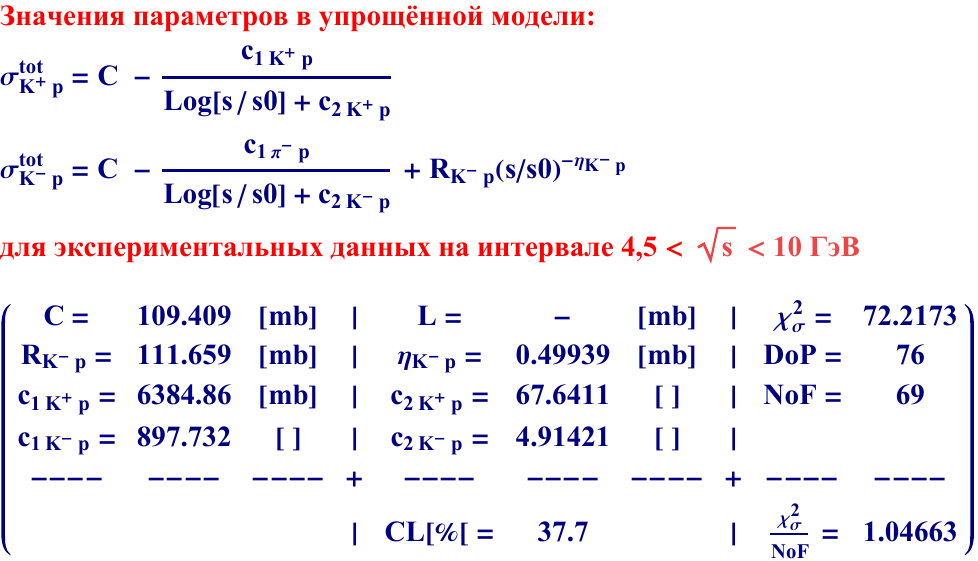}
\vspace{-2.1mm}
\caption{$K^{\pm}p$ reactions and its parameters}
\label{KpPic}
\end{figure}

\newpage

\begin{figure}[h]
~\vspace{-8.1mm}\\
\begin{minipage}[h]{113mm}
\includegraphics[width=113mm]{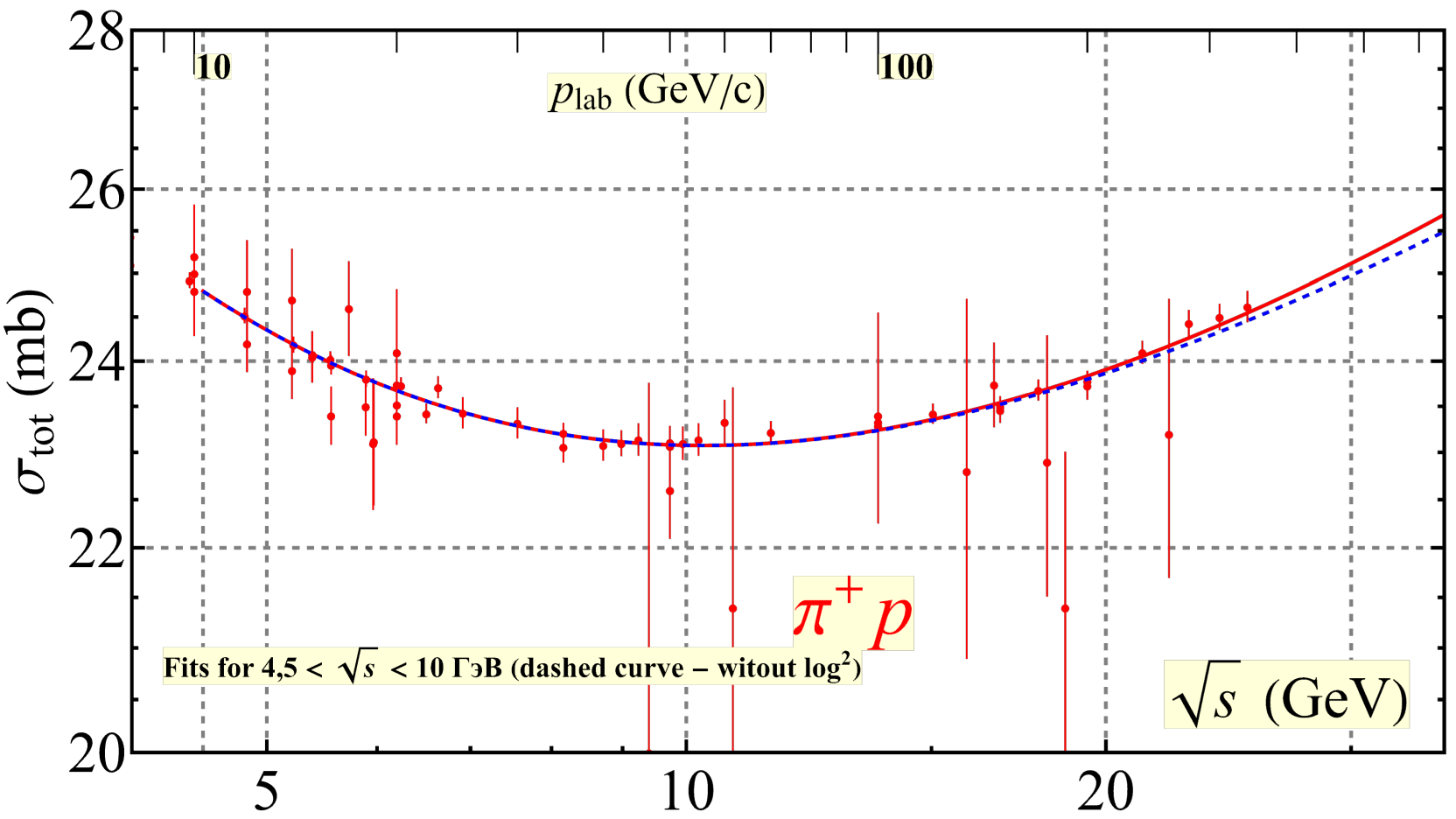}\vspace{-0.1mm}
\end{minipage}
\hfill
\begin{minipage}[h]{59mm}
\includegraphics[width=59mm]{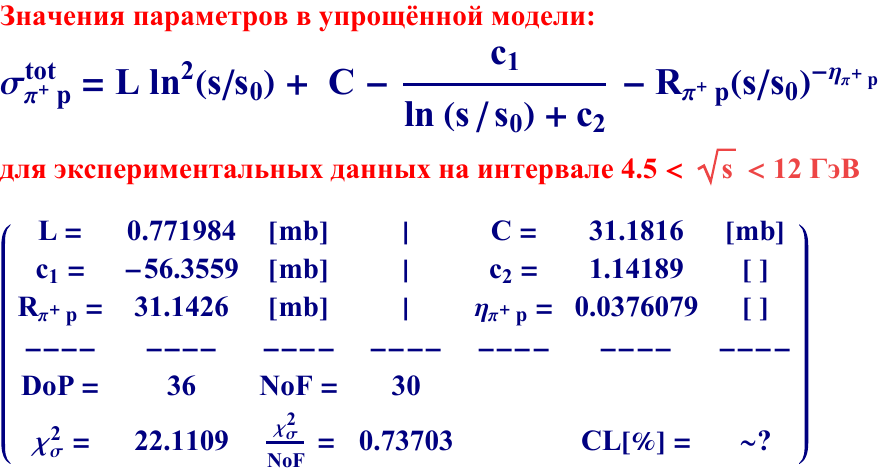}\\
\includegraphics[width=59mm]{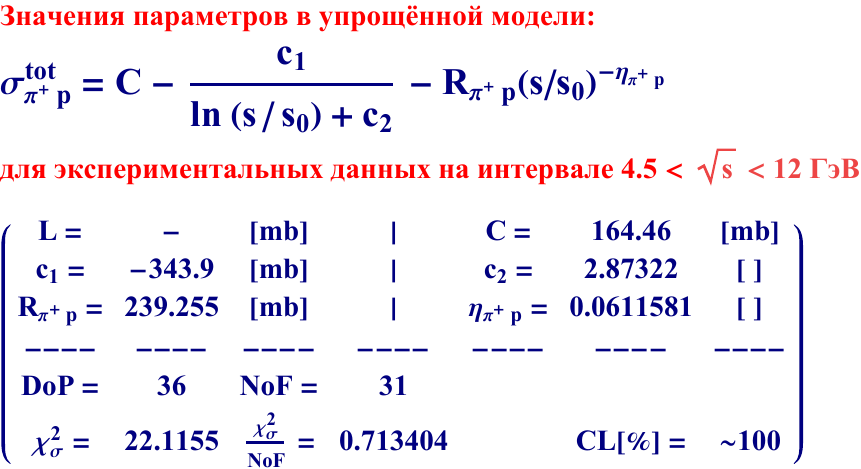}
\end{minipage}
\begin{minipage}[h]{113mm}
\includegraphics[width=113mm]{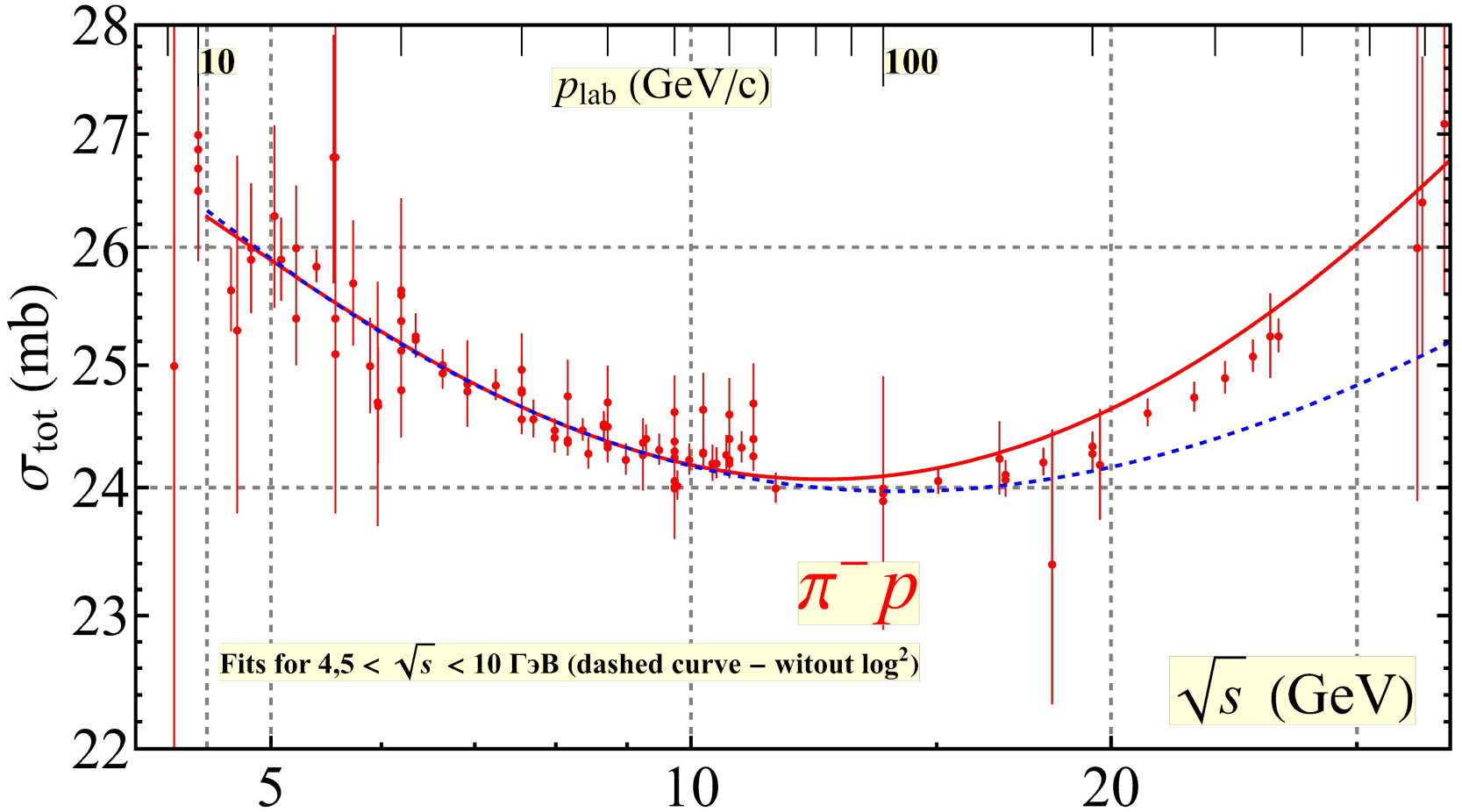}
\end{minipage}
\hfill
\begin{minipage}[h]{59mm}
\includegraphics[width=59mm]{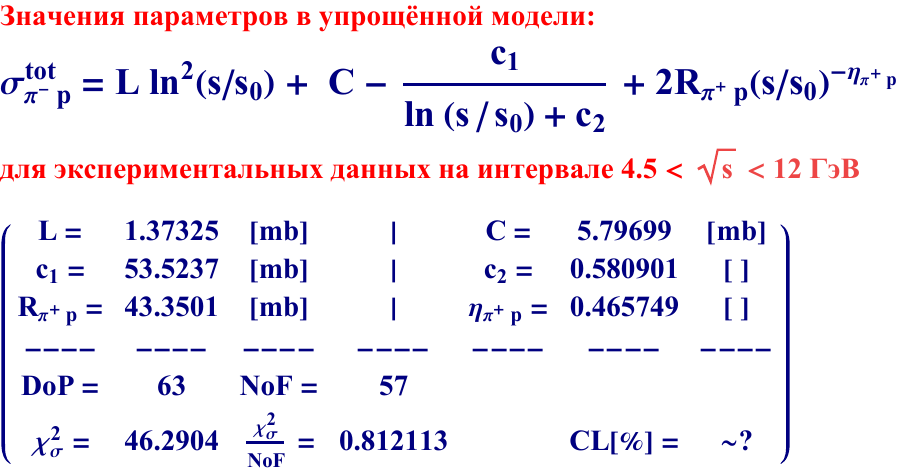}\\
\includegraphics[width=59mm]{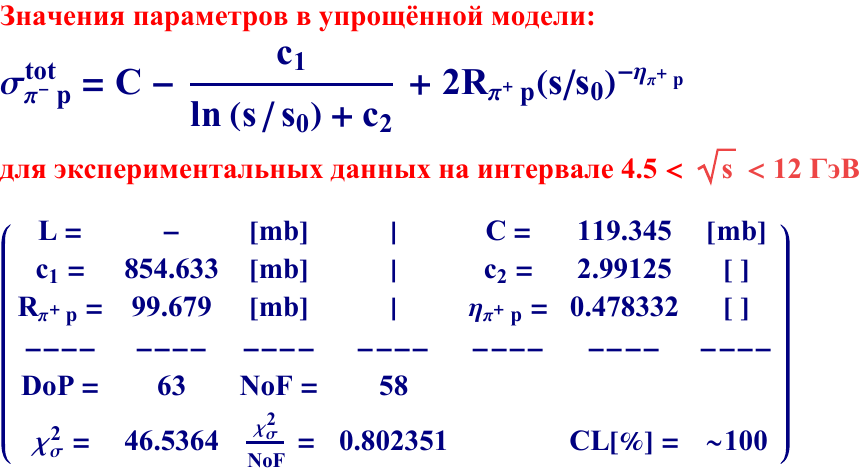}
\end{minipage}
\includegraphics[width=113mm]{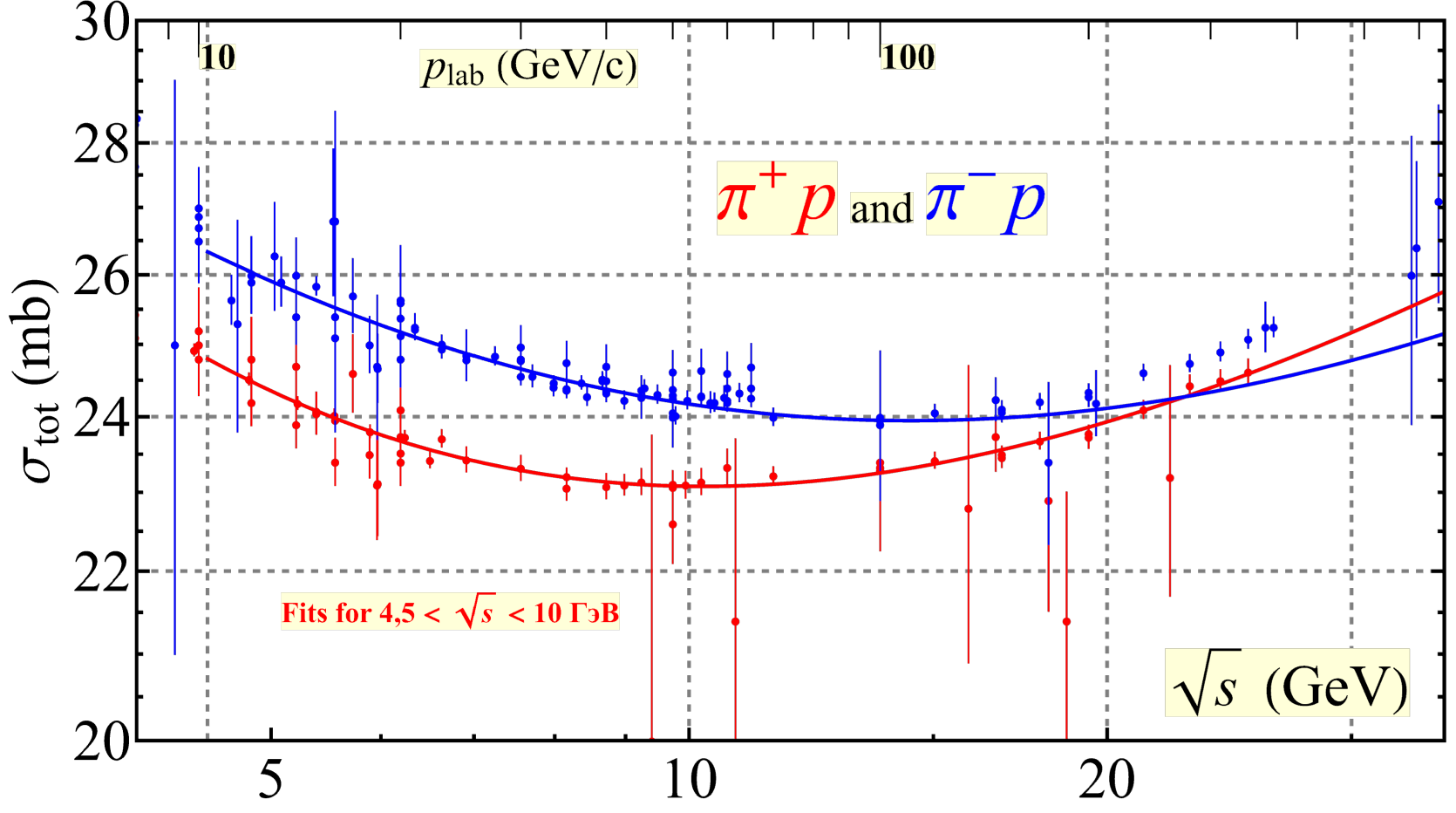} \hfill
\includegraphics[width=60mm]{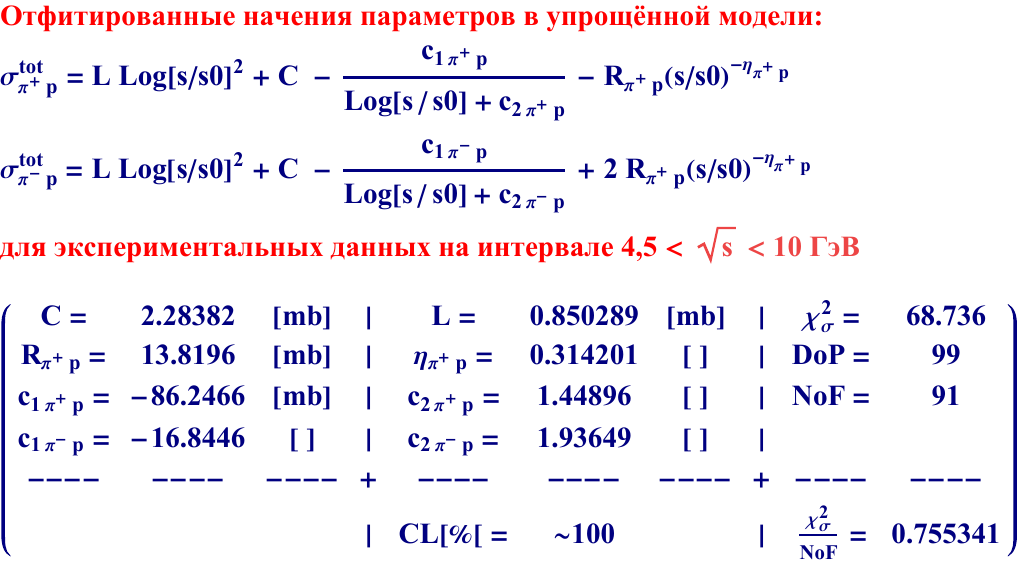}\\
\includegraphics[width=113mm]{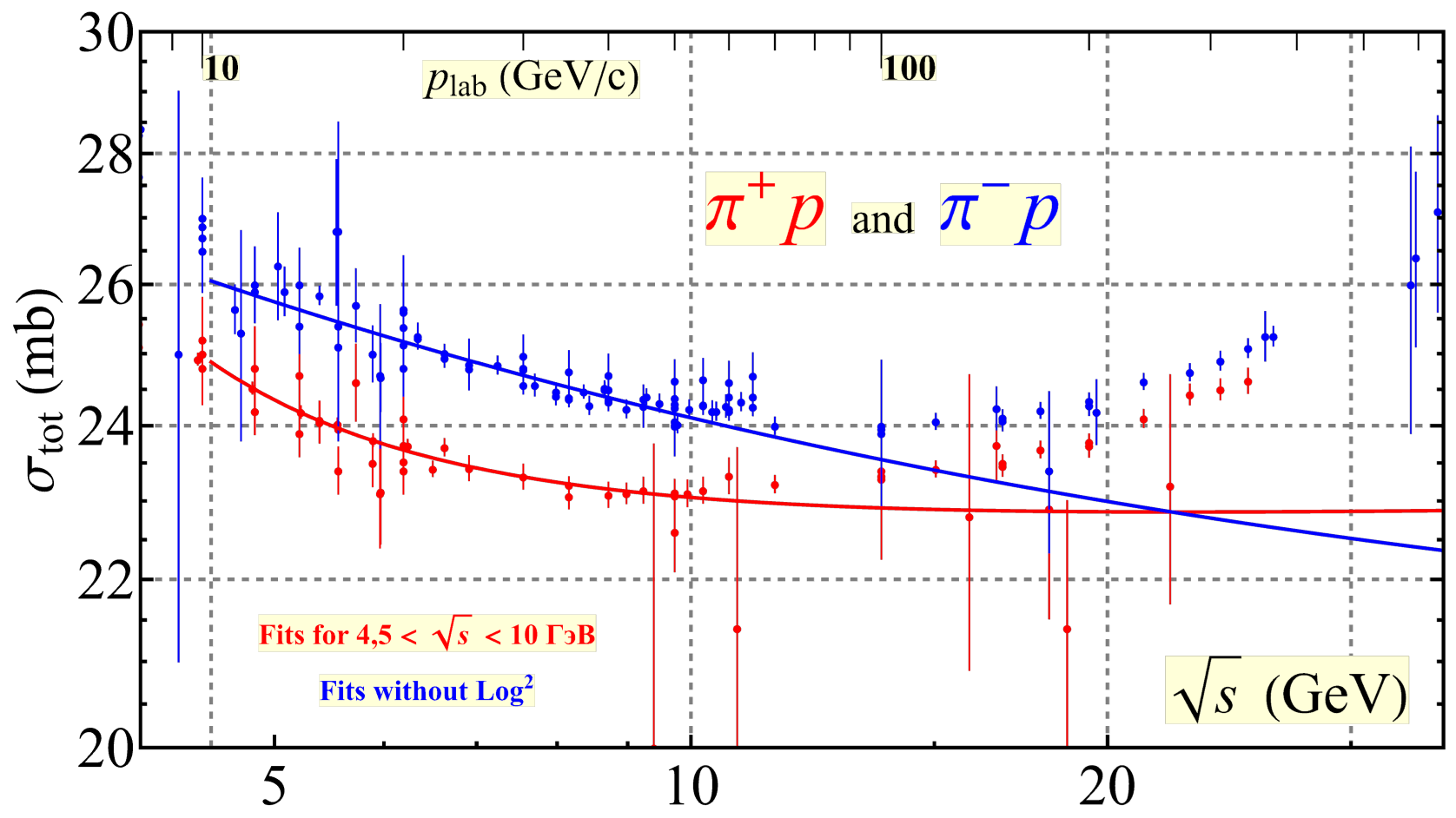} \hfill
\includegraphics[width=60mm]{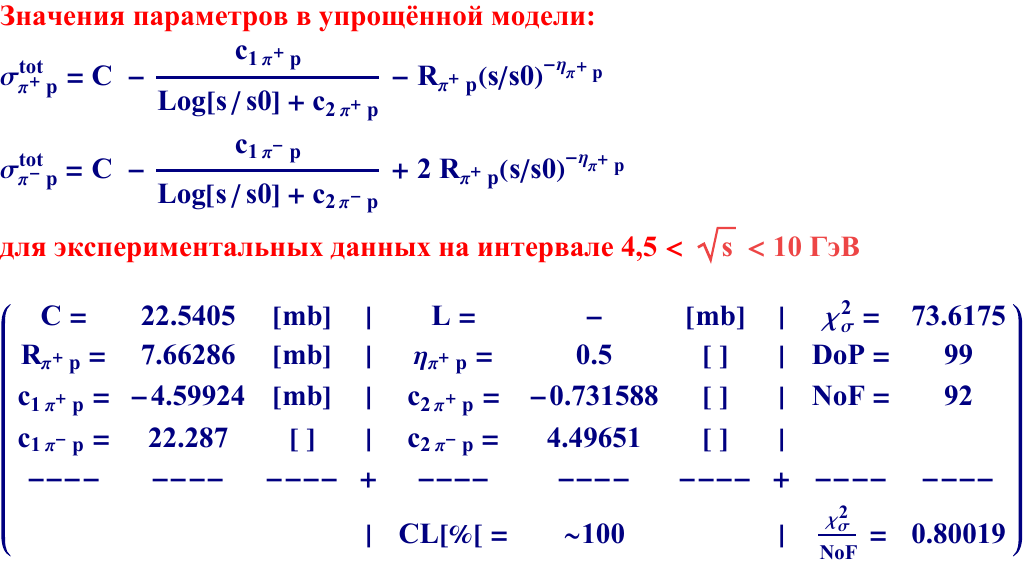}
\vspace{-1.1mm}
\caption{$\pi^{\pm}p$ reactions and its parameters}
\label{PipPic}
\end{figure}


\begin{thebibliography}{99}
\bibitem{Ch1}
G.F. Chew,

The Analytic S-Matrix.

New York: W.A. Benjamin, inc., 1966.

\bibitem{Ch2}
G.F. Chew, Steven C. Frautschi,

Physical Review Letters \textbf{8} (1962) 41.

\bibitem{Pom}
I.Ia. Pomeranchuk,

Soviet Physics, JETP \textbf{7} (1958) 499-501, 
Zh.Eksp.Teor.Fiz.\textbf{ 34} (1958) 725;

\bibitem{Grib}
V.N. Gribov,

Conf.Proc.\textbf{C 670828} (1967) 621; 

V.N. Gribov, I.Ya. Pomeranchuk,

Soviet Physics, JETP \textbf{16} (1963) 220-227, 

Nuclear Physics \textbf{ 38} (1962) 516-528.

\bibitem{Hei}
W. Heisenberg,

Zeitschrift f\"{u}r Physik \textbf{133 }(1952) 65.

\bibitem{Wu}
Hung Cheng, T.T. Wu,

Physical Review Letters \textbf{24} (1970) 1456.

\bibitem {Prok1}
S. P. Denisov et al.,

Physics Letters \textbf{B36} (1971)415;

S. P. Denisov et al.,

Physics Letters \textbf{B36} (1971)528.

\bibitem{Log}
A.A. Logunov, M.A. Mestvirishvili, Nguyen Van Hieu,

Physics Letters B \textbf{25} (1967) 611.

\bibitem {Ezh}
Ezhela V.V., Petrov V.A.,

Preprint IHEP \textbf{72-73}. Serpukhov,1972.
 
\bibitem {Prok2}
Yu.B. Bushnin et al.,

Physics Letters \textbf{B29} (1969)48.

\bibitem{Hei2}
W. Heisenberg. 

Talk at the Niels Bohr Memorial Meeting, Copenhagen, 1963.

See also

W. Heisenberg,

Introduction to the unified field theory of elementary particles.(Eq.(9.5)).

Interscience Publishers, London, New York, Sydney. 1966.

\bibitem {Bar}

V. Bartenev et al.,

Physical Review Letters \textbf{29}(1972)1755. 



\bibitem{Am}

U.Amaldi et al.,

Physics Letters \textbf{B44 }(1973)112;


S.R Amendolia et al.,

Physics Letters \textbf{B44} (1973)119.

\bibitem{QU}
Yu.M. Antipov et al.,

Physics Letters \textbf{B29} (1969) 245.

\bibitem {Ros}
H. Harari,

Physical Review Letters\textbf{22}(1969)562;

J. Rosner,

Physical Review Letters \textbf{22}(1969)689.



\bibitem {Den}
S.P. Denisov, A.V. Kozelov, V.A. Petrov,

Physics of Atomic Nucleus. \textbf{79} (2016) 2, 199.

\bibitem {Ger}
A.S. Gerasimov, A.K. Likhoded, V.A. Petrov, V.D. Samoylenko,

Moscow University Physics Bulletin \textbf{78} (2023) 6, 716,
 
e-Print: 2306.12740 [hep-ph].

\bibitem {pdg}
\href{https://pdg.lbl.gov/2022/hadronic-xsections/hadron.html#xsecplots}{PDG Data Base 2022}

https://pdg.lbl.gov/2022/hadronic-xsections/hadron.html{\#}xsecplots


\bibitem {Tka}

N.P.Tkachenko,

IHEP preprint 90-47. Protvino,1990;

A.G.Abramov et al.,

Cryogenics \textbf{32}(1992)380.

\end{thebibliography}
\end{document}